\documentclass[prb,showpacs,floatfix,twocolumn]{revtex4}
\usepackage{graphicx}
\usepackage{amssymb}
\usepackage{dcolumn}
\usepackage{bm}
\begin{document}

\title{Modeling the buckling and delamination of thin films}
\author{E. A. Jagla}
\affiliation{Centro At\'omico Bariloche, Comisi\'on Nacional de Energ\'{\i}a At\'omica, 
(8400) Bariloche, Argentina}

\begin{abstract}

I study numerically the problem of delamination of a thin film elastically attached to a rigid substrate.
A nominally flat elastic thin film  is modeled using a two-dimensional 
triangular mesh. Both compression and bending rigidities are included to simulate 
compression and bending of the film.
The film can buckle (i.e., abandon its flat configuration) when enough compressive strain is  applied.
The possible buckled configurations of a piece of film with stripe geometry are investigated as a function of
the compressive strain. It is found that the stable configuration  depends strongly on the applied
strain and the Poisson ratio of 
the film. Next, the film is considered to be attached to a rigid substrate
by springs that can break when the detaching force exceeds a threshold value, producing the partial 
delamination of the film.
Delamination is induced by a mismatch
of the relaxed configurations of film and substrate. 
The morphology of the delaminated film can be followed and compared with available experimental 
results as a function of model parameters. 
`Telephone-cord', polygonal, and `brain-like' patterns qualitatively similar to experimentally observed configurations
are obtained in different parameter regions. The main control parameters that select the different 
patterns are the mismatch between film and substrate and the degree of in-plane relaxation within the unbuckled regions.

\end{abstract}
\maketitle

\section{Introduction}

The use of a variety of coatings to enhance the performance of materials is
widespread in many areas of science and technology. The deposition of the
coating usually occurs at conditions (such as temperature, humidity of the environment, etc) very
different of those to be found under work conditions. Due to this fact and to the different nature of
substrate and coating, typically large mismatch
stresses appear between the film and the substrate. When the stresses are tensile on the film, this may 
fail duo the the nucleation of cracks that split the film to reduce 
to total mechanical energy of the system. A familiar example of this phenomenon
is the cracking observed sometimes in paints and in the surface of mud.\cite{barro} 
When mismatch stresses are compressive within the film, the
most common mode of failure is called delamination:\cite{delamination1,delamination2,delamination3} 
the film  partially detaches from the substrate to relieve 
the accumulated stress. Experimentally, different morphologies of the delaminated regions have been
observed. In many cases delaminated regions of characteristic undulated geometries (referred to  as 
`telephone cords') appear.

Delamination can be understood on the basis of the well known phenomenon of buckling of elastic 
structures.\cite{timosh}
The first quantitative description of the buckling 
of a one dimensional elastic rod goes back to Euler. Upon compression, an elastic rod can reach 
a state in which the straight configuration is no longer the minimum energy configuration. 
At that point the rod acquires a non-straight configuration, namely, it buckles. For two 
dimensional elastic membranes the same kind of instability exists, but the problem becomes 
mathematically much more involved. The equilibrium equations of a membrane were obtained 
by Foppl and von Karman (FvK).\cite{fvk} The FvK equations are two non-linear, coupled differential 
equations for the separation of a given point with respect to the flat configuration, and 
for the Airy potential for the in-plane displacement. The in-plane displacement can be
reconstructed from auxiliary (linear) differential equations once the Airy potential is 
known.

Delamination of a film that is elastically attached to a rigid substrate involves the 
interplay of two main ingredients. For a given initial form of 
a delaminated region (in which the interaction with the substrate is assumed to be absent) the FvK 
equations determine the geometry of the buckled configuration. The buckled film produces forces on the border
that tend to detach the film further. In this way the film can continue its detaching by this mechanism, 
with the configuration of the buckled film adapting to the instantaneous form of the buckled region.
Thus, in general, buckling and delamination of the film must be solved self-consistently, and this is possible
only numerically. 

\begin{figure}[h]
\includegraphics[width=8cm,clip=true]{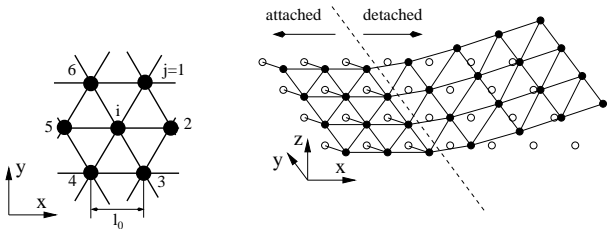}
\caption{(a) Detail of the triangular lattice that simulates the film, where the numbering of neighbors of site $i$ and the equilibrium
distance $l_0$ are highlighted. (b) A sketch of part of a partially delaminated film. Open symbols are the anchoring points in the 
substrate. The film is attached to the substrate to the left, and
is detached to the right. Note the horizontal relaxation in the attached part, indicated by the thin lines, along which the spring constant
$k_h$ is acting.}
\label{f1}
\end{figure}

A direct numerical solution of the FvK equations is rather difficult, because 
of its non-linear nature. This is why some people have considered the simulation of a more-or-less 
realistic film, consisting of atoms interacting via inter-atomic forces. The problem with this approach is 
that in order to be able to model the bending of the film, a few layers of atoms are necessary, making the
simulations rather inefficient, and suffering of important spurious effects associated to the 
anisotropy of the numerical lattice.\cite{simulaciones2}
Here, to simulate the film I use the following strategy, that to a large extent eliminates the
previously mentioned drawbacks.\cite{nota}
The film is modeled on a physical basis by a triangular lattice formed by springs that account for
the compression/stretching behavior of the material. In addition, bending springs are introduced in the
sites of the triangular lattice that account for the bending rigidity of the film. The model can be 
shown to be elastically isotropic (in linear approximation), describing a material with Poisson ratio $\nu=1/3$. 
Additional bending springs are added to change the value of $\nu$ in a controlled manner, 
keeping the system isotropic. 
By referring to 
the sketch in Fig. \ref{f1}(a), each node $i$ of the triangular lattice contributes with an elastic energy $H_i$ of the form:

{\widetext

\begin{equation}
H_i=
\sum_{j=1}^6 \frac 14 k_s\left [ |{\bf u}_{ij}|-l_0  \right]^2+\sum_{j=1}^3  k_b\left ( 
\frac {{\bf u}_{ij} . {\bf u}_{ij'}}{|{\bf u}_{ij}||{\bf u}_{ij'}|} +1  \right)
+\sum_{j=1}^6\frac 12 k_{\nu}\left ( 
\frac {{\bf u}_{ij} . {\bf u}_{ij''}}{|{\bf u}_{ij}||{\bf u}_{ij''}|} -\frac 12  \right)^2
\label{energia}
\end{equation}}
\endwidetext
where ${\bf u}_{ij}$ is the vector joining sites $i$ and $j$, the index $j$ labels the six neighbors of 
site $i$, in sequential order, $j'$ in the second term is $j+3$ (i.e, the neighbor 
opposite to neighbor $j$), and $j''$ in the last term is $j''=j+1$ for $j$ from 1 to 5, and $j'=1$ for $j=6$. 
The three terms of this expression are respectively the stretching energy, the bending energy, and the corrective term
that is introduced if a film with $\nu\ne 1/3$ is required. They are characterized by three different elastic constants, 
namely $k_s$, $k_b$, and $k_\nu$. 
The equilibrium configuration of a system described by the energy (\ref{energia}) is a flat triangular lattice with 
lattice parameter $l_0$.

From this atomistic expression for the energy of the system, the relation with averaged macroscopic 
quantities can be evaluated. The result for the bending rigidity $D$, the two-dimensional 
compressibility $B$ and the Poisson ratio $\nu$ of the bulk material of the film (which is assumed to have a
thickness $d$) is\cite{ojo}
\begin{eqnarray}
B=\frac{{\sqrt 3}}{d}k_s\nonumber\\
D=\frac{3\sqrt3}{4}k_b\label{k-e}\\
\nu=\frac{2l_0^2k_s-9{k_\nu}}{6l_0^2k_s+9{k_\nu}}\nonumber
\end{eqnarray}

The attaching (and eventual detaching) of the film to the substrate is the second crucial ingredient
to include in order to obtain a realistic behavior of the system. One possibility (as used for instance 
in Ref. \cite{simulaciones2}) is to consider that when attached to the substrate the film is rigidly pinned to it (i.e., it cannot
move at all)
whereas, if the
force exerted by the substrate onto the film exceeds some threshold value, the interaction to the substrate is set 
irreversible to zero. Experimentally, it has been emphasized that detaching by perpendicular
separation has usually a much lower threshold
than detaching by shear.\cite{delamination3} This suggests that an appropriate situation is to consider that detach occurs when the perpendicular
force between film and substrate exceeds some maximum value $f_{max}$. This was the scheme adopted in \cite{simulaciones2}. We will 
see that this prescription does not suffice to reproduce the large phenomenology that is observed experimentally.
I think that the main lacking ingredient is the possibility of in-plane relaxation of the film in the case it is still attached
to the substrate. This relaxation is introduced here in the energetics of the model in the following way (see Fig. \ref{f1}(b)). 
Let us consider the substrate to be located at $z=0$. For each point of the film there is an anchoring point on the substrate to 
which this is attached to. The coordinates of the anchoring point are ($x_0$, $y_0$, 0). The coordinates of the corresponding point in 
the film are ($u_x$, $u_y$, 0). The contribution to the energy from this link is $\frac 12 k_h [(u_x-x_0)^2+(u_y-y_0)^2]$. 
The value of $k_h$ determines the extent of in-plane relaxation. For $k_h\rightarrow\infty$ we recover 
the rigid attaching condition. 
When the
$z$ force that is necessary to apply to a given point in the film to keep it at zero $z$ coordinate overpasses the threshold value $f_{max}$, the value
of the corresponding $k_h$ is set irreversibly to zero, and the $z$ coordinate is not any more restricted to be zero.
The anchoring points in the substrate form a triangular lattice, corresponding to that of the film. However the lattice parameter of this
lattice $l_0^s$ is assumed to be smaller than the one $l_0$ in the film. This is the way in which a mismatch between film and substrate appears,
and the origin of the tendency of the film to delaminate.
Macroscopically, the attaching to the substrate can be quantified by an elastic coefficient we
call $E_s$, that is defined as the ratio between the applied horizontal force per unit area of the film and the horizontal displacement 
this force produces. The relation between $k_h$ and $E_s$ is easily found to be
\begin{equation}
E_s=\frac{2k_h}{\sqrt 3l_0^2}
\end{equation}

In addition to the energetics, the time evolution of the system must be defined. 
This evolution will be assumed to be overdamped, namely it is assumed that kinetic energy
plays no role in the evolution. Then the positions of the nodes of the lattice evolve according to:

\begin{equation}
\frac{d{\bf u}}{dt}=-\lambda \frac {\delta H}{\delta {\bf u}}
\label{dinamica}
\end{equation}
thus $\lambda$ sets the time scale for the evolution.

It is also convenient to introduce non-dimensional parameters to describe the simulations. From (\ref{energia})
and (\ref{dinamica}), it can be seen that the simulations can be parametrized by the values of $k_b/(l_0^2k_s)$, 
$k_{\nu}/(l_0^2k_s)$, and $f_{max}/(k_sl_0)$. A non-dimensional time $\tau$ can also be introduced through
$\tau=\lambda k_s t$. All simulations presented below where made in systems with periodic boundary conditions, using a first
order method to integrate (\ref{dinamica}), with a time step $\delta\tau=0.3$.

\section{Buckling patterns of a stripe}

Before studying the full interplay between buckling and delamination, it is convenient to consider
the buckling patterns obtained for a fixed delaminated region. I present the case of a delaminated
region with stripe geometry because this case has been considered previously in the literature,\cite{audoly}  and then it 
will be useful to validate the numerical model. In addition we will see that already in this simple case the 
behavior is non-trivial.
Thus we consider a stripe piece of the film of width $w$, that is detached from the substrate. The rest of the film is assumed
to be rigidly attached to the substrate (i.e., $k_h\rightarrow\infty$). In this way, the only effect of this part of the film is to provide
a clamped boundary condition for the stripe. In addition, due to the presence of the substrate, the stripe can buckle only toward
positive values of $z$. Under these conditions we will look for the stable configuration of the stripe for different parameters.

We define the mismatch strain between film and substrate $\varepsilon$ as $\varepsilon\equiv (1-l_0^s/l_0)$.
When the stripe is loaded isotropically in compression (i.e, $\varepsilon>0$), the film remains flat until a 
critical value  $\varepsilon_{1}$ is reached. At this point
a uniform wrinkle along the longitudinal direction is formed. This configuration is usually refer to as the Euler column. 
The value of $\varepsilon_{1}$ can be analytically calculated, and it is known to be\cite{audoly}
\begin{equation}
\varepsilon_{1}=\frac{4\pi^2D}{w^2Bd}
\end{equation}
An equivalent expression can be obtained by expressing $D$ and $B$ in terms of the Young modulus of the material $E$, es indicated in \cite{ojo}.
The result is
\begin{equation}
\varepsilon_{1}=\frac{\pi^2 d^2}{3w^2(1+\nu)}
\end{equation}
where we see that the value of $\varepsilon_1$ is mainly controlled by the ratio between thickness and width of the delaminated stripe.

The value of $\varepsilon_1$ can also be expressed in terms of the parameters of the model by using (\ref{k-e}), as

\begin{equation}
\varepsilon_{1}=\frac{4\pi^2}{n^2}\frac{k_b}{l_0^2k_s}
\end{equation}
where $n$ is the number of rows of particles that compose the delaminated stripe (i.e., $w=\frac{\sqrt{3}}{2}l_0n$), and where the previously
defined non-dimensional parameter $\frac{k_b}{l_0^2k_s}$ appears.

The Euler column configuration mostly relieves the perpendicular stress on the stripe, but leaves
an important degree of longitudinal compression in it. If the isotropic compression $\varepsilon$ is increased,
a secondary buckling is expected at some mismatch strain $\varepsilon_2$, in which the longitudinal uniformity of the Euler column is lost, and one 
of two different patterns is expected:\cite{audoly}
in one case (occurring if the Poisson ratio $\nu$ of the film is larger than a value of approximately 0.255) the Euler column transforms
in an undulated wrinkle. The second case corresponds to a pattern (called sometimes `varicose')
appearing if $\nu$ is lower than this value, in which the uniform maximum of the Euler column is distorted and transforms in a sequence of small hills
and shallow valleys. The varicose pattern preserves the symmetry of the configuration with respect to the axis of the delaminated region, whereas the 
undulated wrinkle does not. These results have been obtained analytically doing a perturbative
analysis of the FvK equations considering small distortions of the Euler column. The analysis shows also\cite{audoly}
that $\varepsilon_2/\varepsilon_{1}$ is a function of the Poisson ratio of the film only. 
Using the numerical model presented here, we can test this prediction, and in addition we can explore arbitrary distortions of the
film. 

\begin{figure}
\includegraphics[width=8cm,clip=true]{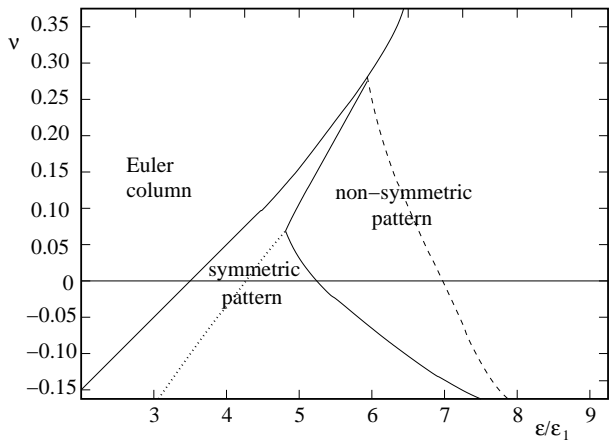}
\caption{The $\nu$-$\varepsilon$ parameter space, with the different sectors in which different qualitative behavior is observed. The
lines separating different regimes were obtained by doing independent simulation at different point of the $\varepsilon$-$\nu$ plane. Some of
them are presented in the next three figures.
See the text and Figures \ref{f3}, \ref{f4}, and \ref{f5}.
Simulations were performed using a triangular lattice of 200 sites along the stripe, and 24 rows in the perpendicular direction. The 
dimensionless combination of parameters $k_b/(k_s l_0^2)$ was taken to be 0.05. The value of the critical strain $\varepsilon_1$ is 
in this case $\varepsilon_1\simeq 0.0032$.
}
\label{f2}
\end{figure}

\begin{figure}
\includegraphics[width=8cm,clip=true]{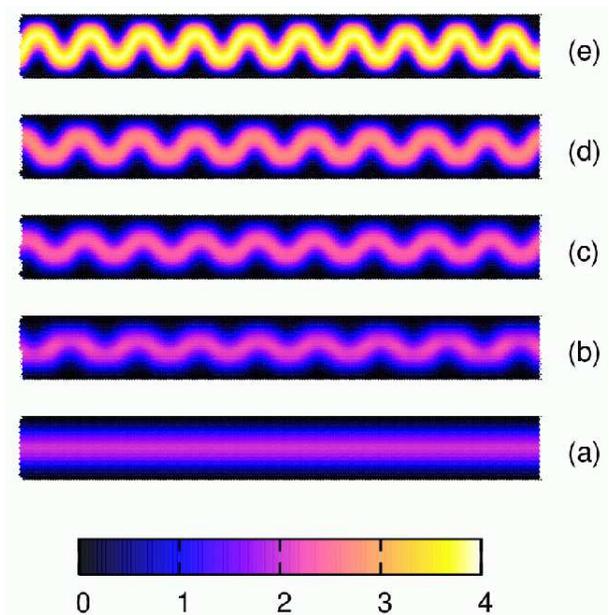}
\caption{(Color online) The buckled configuration of the stripe, for $\nu=1/3$, and $\varepsilon/\varepsilon_1= 6.25$, 7.5, 9.4, 12.5, and 25, 
respectively form
(a) to (e). The gray (color) scale indicates the departure of the film from the substrate, in units of the discretization distance $l_0$.
}
\label{f3}
\end{figure}

\begin{figure}
\includegraphics[width=8cm,clip=true]{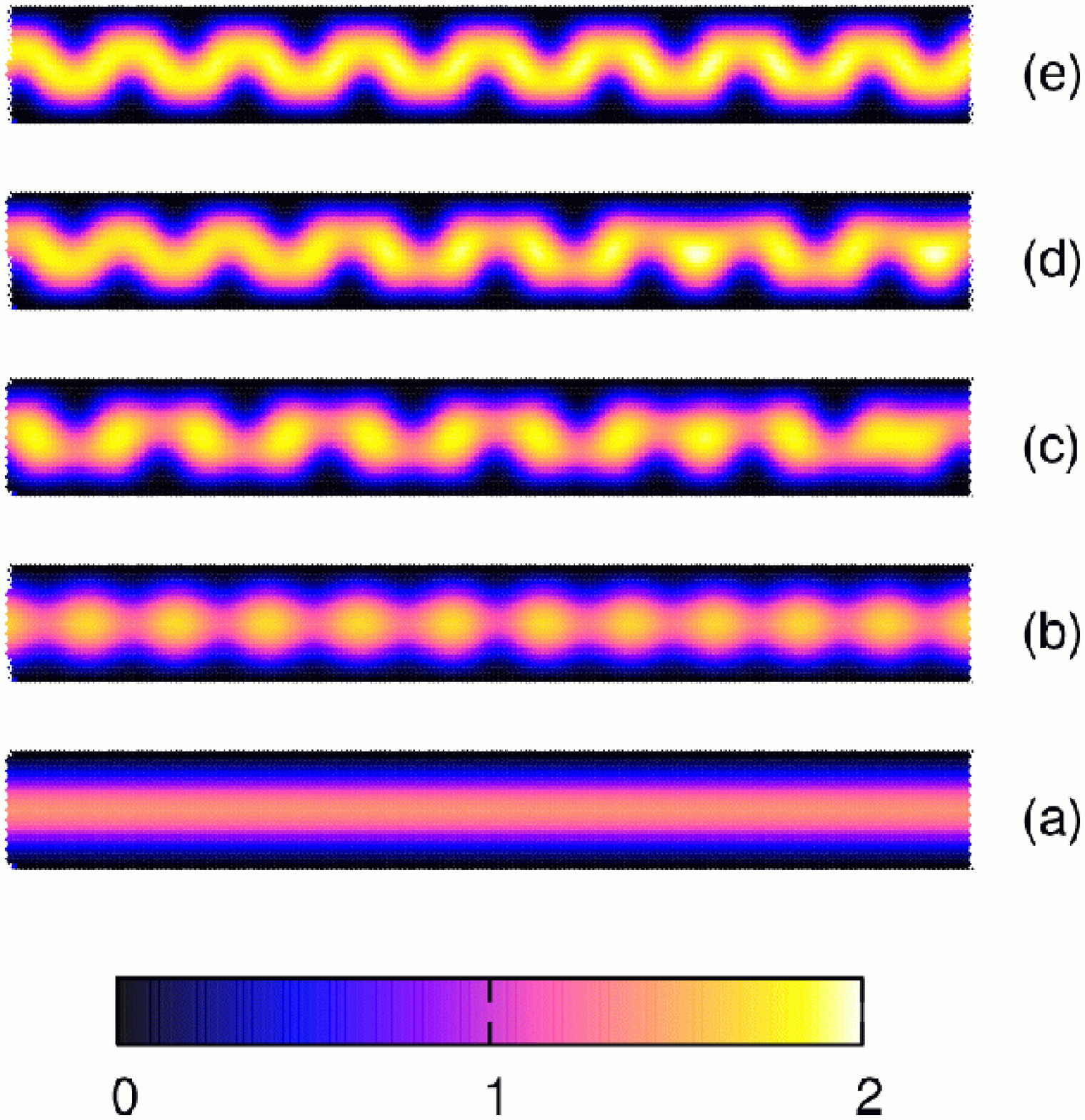}
\caption{(Color online) Same as Fig. \ref{f3} for $\nu=0.075$, and $\varepsilon/\varepsilon_1= 4.0$, 4.7, 5.6, 6.25, and 6.9, respectively form
(a) to (e).
}
\label{f4}
\end{figure}

\begin{figure}
\includegraphics[width=8cm,clip=true]{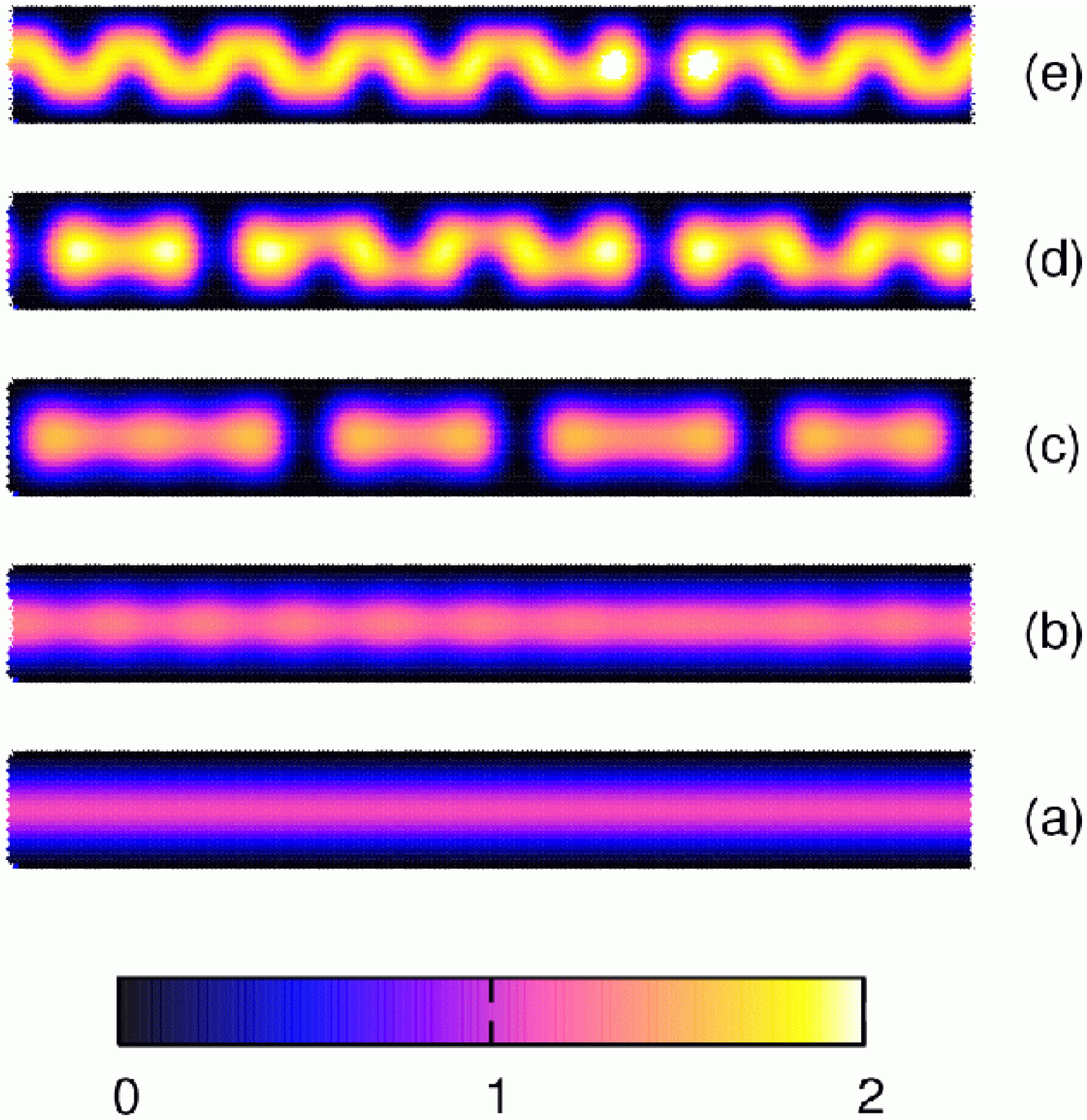}
\caption{(Color online) Same as Fig. \ref{f3} for $\nu=0$, and $\varepsilon/\varepsilon_1= 3.12$, 3.75, 4.37, 6.25, and 6.9, 
respectively from (a) to (e).
}
\label{f5}
\end{figure}

Numerical results in the stripe geometry were obtained taking one of the main directions of the triangular simulation lattice
along the stripe. The stripe was described by 24 rows of this triangular mesh, and the value of $k_b/(l_0^2k_s)$ was taken to be 0.05.\cite{nota2}
Results are synthesized in the phase diagram of Fig. \ref{f2}, and in Figs. \ref{f3}, \ref{f4}, and \ref{f5}, 
where the obtained configurations are shown for different values of the 
two main control parameters $\varepsilon/\varepsilon_1$ and $\nu$.
In agreement with the analytical predictions, we can see  that by increasing $\varepsilon$, the  instability of the Euler column corresponds
to an non-symmetric pattern for $\nu>\nu_c\simeq 0.25$, whereas it corresponds to a symmetric one for $\nu<\nu_c$. However, we also see
that when larger values of $\varepsilon$ are considered, the stability range of the symmetric pattern is (at least for $\nu \gtrsim 0.05$) only a rather
small wedge in the phase diagram. For sufficiently large values of $\varepsilon$ the undulated wrinkle pattern is observed 
always to be the stable one.

The symmetric and non-symmetric sectors of the phase diagram in Fig. \ref{f2} have been divided by a dotted and dashed line, respectively. 
I describe now their meaning.
The dashed line in the non-symmetric sector marks only a crossover: At the right of this line, the undulated wrinkle is well
develop, in the sense that it has a maximum of almost constant amplitude that undulates along the stripe, see for instance Fig. \ref{f4}(e).
At the left of the dashed lined, there is a clear modulation of the amplitude (Fig. \ref{f4}(c)). I emphasize again that the transition between the two different behaviors
is simply a crossover.

The dotted line within the symmetric sector in Fig. \ref{f2} marks an interesting behavior change in the region of small or negative
values of $\nu$. 
To the left of the dotted line we have the symmetric pattern as described analytically in Ref. \cite{audoly}.
To the right of the dashed line, this pattern gets modified by the appearance
of deep valleys interrupting the normal sequence of small hills and 
shallow valleys (see Fig. \ref{f5}(c)). Note that in the center of the deep valleys, the film is almost touching the substrate. 
The appearance of these deep valleys is abrupt, namely they do not grow deeper and deeper as some parameter is changed, but they simply 
exist or do not exist. 
The periodicity of the appearance of deep valleys is difficult to obtain from the simulations, but the information collected from different
simulations suggests that close to the dotted line of the phase diagram deep valleys are well separated from each other, 
whereas when moving to larger values of $\varepsilon$ they become closer, 
eventually leaving no shallow valleys in between.
A qualitative explanation for the appearance of these deep valleys could be based on the following observation. 
In a material with negative Poisson ratio, a compression in one
direction is accompanied by a compression in the perpendicular direction. Transforming a shallow valley of a varicose pattern into a deep valley
produces a perpendicular compression of the film, that implies a tendency to compress also in the longitudinal direction. In this way
part of the longitudinal compression is relieved, and this can stabilize the deep valley. 

The previous results show that, for the usual cases of materials with positive Poisson ratio close to 1/3, the undulated wrinkle pattern is the
most likely to be observable in experimental situations in which stripe delaminated regions occur.

\section{Delamination}

In this Section I consider the full interplay of buckling and delamination, and present the different kinds of delamination 
morphologies that can be obtained with the present model.
From now on I restrict to the case of a film with $\nu=1/3$ (i.e., $k_{\nu}=0$). When buckling and delamination
both occur, there are two natural length 
parameters in the problem whose meaning is convenient
to emphasize from the beginning. 
The first length scale $L_1$ is the typical size of a blister that is in equilibrium upon further delamination. AN estimation of $L_1$ goes
as follows.
For a buckled region of typical linear size $L_1$ ($L_1$ may be the for instance 
the diameter of a circular blister, or the width of an Euler column) under a deformation $\varepsilon$, the typical curvature of the 
buckled film at the points of contact with the substrate is $\sim  \sqrt{ \varepsilon}/L_1$.\cite{otro} 
The torque $T$ per unit length 
exerted by the delaminated film on the part attached to the substrate is obtained by multiplying this curvature
by the bending rigidity $D$, i.e, $T=\sqrt{ \varepsilon} D/L_1$. 
The equilibrium condition for the film will correspond to this torque been equal to a critical maximum value that the 
interaction with the substrate can sustain. I note this maximum torque as $T_{max}$. Thus we obtain the order of magnitude of $L_1$ as 
$L_1\sim \sqrt{ \varepsilon} D/T_{max}$. It is important to emphasize that the condition of a maximum torque per unit length that I use here
is totally compatible with the condition of a maximum force used to describe the model in the previous Section. In fact, note first
of all that a torque per unit length has the correct units of a force. In addition, a detailed consideration of the geometry
of the triangular lattice used, 
shows that $T_{max}$ in the continuum description and $f_{max}$ of the atomistic
description are simply related by $T_{max}=f_{max}\sqrt{3}/2$. Then, in the
microscopic units of the model $L_1$ can be written also as $L_1 \sim \sqrt{\varepsilon} k_b/f_{max}$.

\begin{figure}
\includegraphics[width=8cm,clip=true]{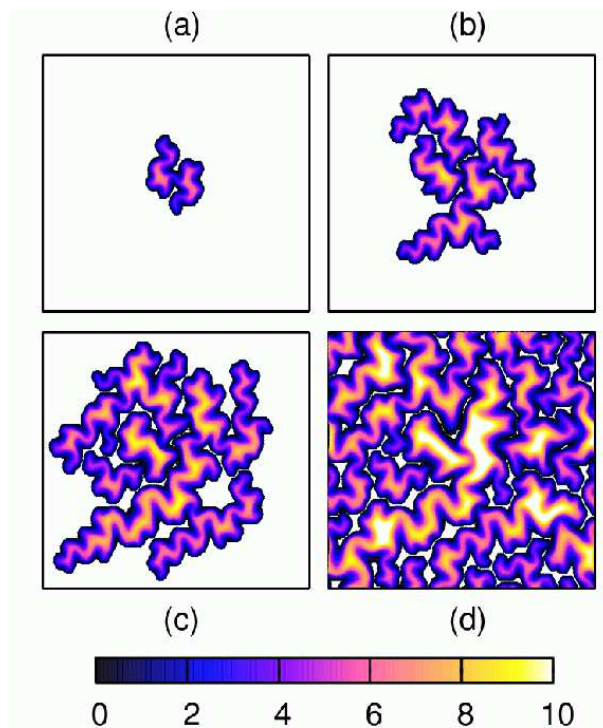}
\caption{(Color online) Temporal evolution of the delaminated regions for $\varepsilon=0.09$, $k_b/(k_sl_0^2)=0.05$, 
$f_{max}/(k_sl_0)=0.03$, and $k_h=k_s$.
Note that the gray (color) scale goes from darker to brighter when the vertical coordinate of the film increases, but
non-delaminated regions are plotted as white for clarity. Vertical scale is in units of $l_0$. Dimensionless time $\tau$ for the four views are 
$\tau=5\times 10^3$, $2\times 10^4$, $3.5\times 10^4$, and $2\times 10^5$.
}
\label{f6}
\end{figure}

\begin{figure}
\includegraphics[width=8cm,clip=true]{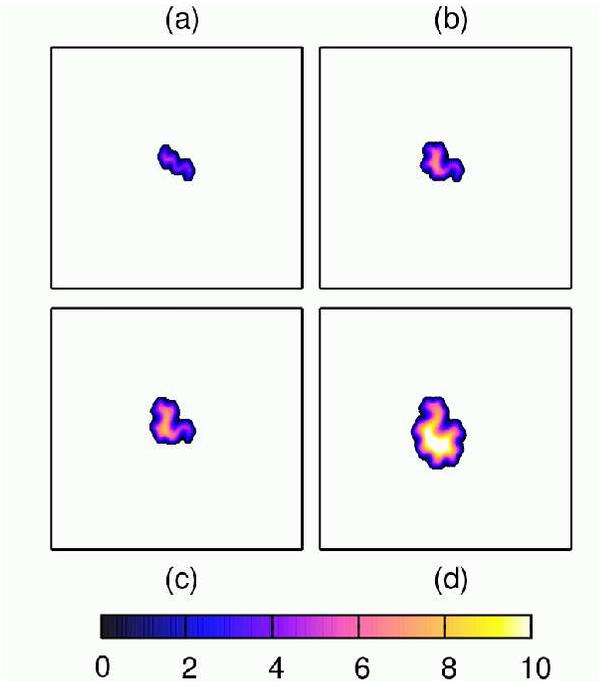}
\caption{(Color online) Same as Fig. \ref{f6} for $\varepsilon=0.085$,  and $k_h=k_s$, at times 
$\tau=5\times 10^3$, $2\times 10^4$, $3.5\times 10^4$, and $2\times 10^5$ . Note in particular that the
last panel shows a stable configuration in which the film does not delaminate further.
}
\label{f7}
\end{figure}

\begin{figure}
\includegraphics[width=8cm,clip=true]{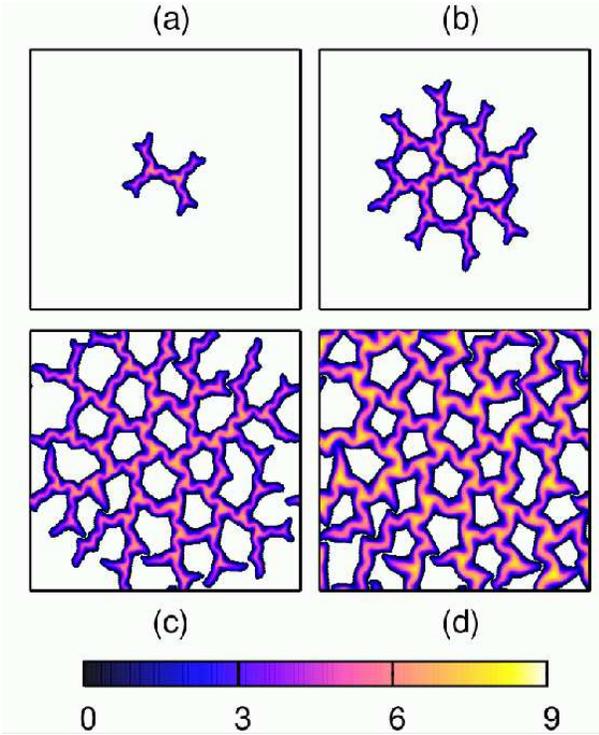}
\caption{(Color online) Same as Fig. \ref{f6} for $\varepsilon=0.09$, $k_h=k_s/100$, 
at times $\tau=1.2\times 10^3$, $3\times 10^3$, $5.4\times 10^3$, and
$1.8\times 10^4$.
}
\label{f8}
\end{figure}

\begin{figure}
\includegraphics[width=8cm,clip=true]{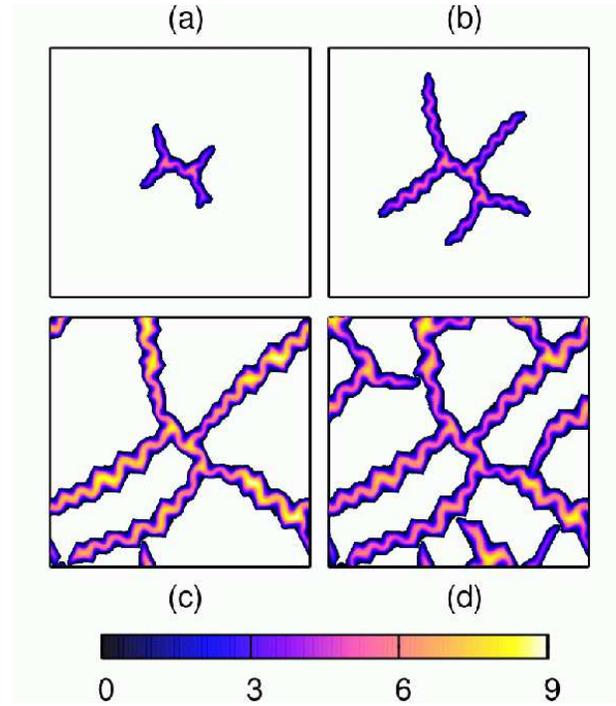}
\caption{(Color online) Same as Fig. \ref{f6} for $\varepsilon=0.07$,  $k_h=k_s/100$, at times $\tau=
2.4\times 10^3$, $5.4\times 10^3$, $3.5\times 10^4$, and $1\times 10^5$.
}
\label{f9}
\end{figure}

\begin{figure}
\includegraphics[width=8cm,clip=true]{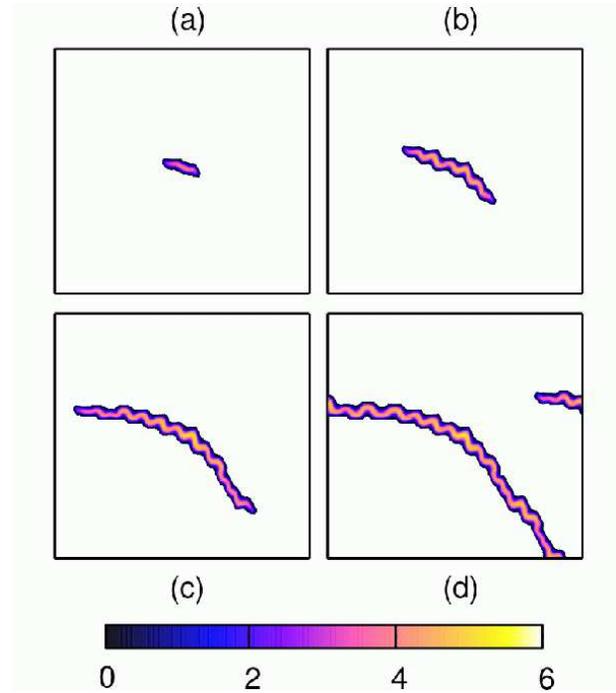}
\caption{(Color online) Same as Fig. \ref{f6} for $\varepsilon=0.065$, $k_h=k_s/100$, at times $\tau=
6\times 10^2$, $3.6\times 10^3$, $8.4\times 10^3$, and $1.4\times 10^4$.
The curvature observed
is due to a residual effect of the triangular lattice that defines the model.
}
\label{f10}
\end{figure}

A second characteristic distance is related to the existence of horizontal relaxation within the unbuckled film.
In fact, consider a film uniformly attached to the substrate. If we force a horizontal displacement
of some point of the film, this perturbation will decay away in a typical distance $L_2$, which depends on the ratio between
the stiffness of the film and the strength of the interaction to the substrate, namely $L_2\sim l_0 \sqrt{k_s/k_h}$, or in macroscopic units
$L_2\sim \sqrt{Bd/E_s}$.
The ratio $L_2/L_1$ is a measure of the importance of horizontal relaxation. This will be negligible if $L_2/L_1 \ll 1$, or important
if $L_2/L_1 \gtrsim 1$.

In order to observe the delamination process, I seed the simulation with a configuration
in which a small part of the film is detached. This portion was chosen as a small rectangular piece, mis-oriented with respect to the numerical mesh to
minimize spurious effects associated to it. In a first stage of the simulation I take $f_{max}\rightarrow \infty$. In this way 
buckling of the 
seeded delaminated region occurs, but the film does not delaminate further. After a stationary configuration is achieved, 
$f_{max}$ is put to a finite value, time is reset to zero, and the evolution of the film is calculated. In all cases presented the numerical lattice
has a total of 300 $\times$ 300 nodes, and periodic boundary conditions are used.

I present first some result in the case the in which horizontal relaxation is small, namely  $k_h=k_s$. This corresponds to a distance $L_2$
of the order of the discretization distance $l_0$, i.e., a very small value. 
In fig. \ref{f6} we see a sequence of configurations corresponding to $\varepsilon=0.09$. We see how in this case the film delaminates almost
completely. This kind of pattern was called `brain-like' in \cite{delamination2}. In fig. \ref{f7} we see the situation for a slightly 
lower value of $\varepsilon$, namely $\varepsilon=0.085$.
After the delamination of a small region near the original defect, the evolution stopped completely, and delamination halted. 
We see that there is a rather sharp transition between a blister that is not able to grow for low strains, to an almost
completely delaminated film for larger strains. 

In the case in which the value of horizontal relaxation is higher, there are important differences in morphology. Figs. \ref{f8}, \ref{f9},
and \ref{f10} show the evolution of the same original defect for the case of $k_h=k_s/100$ (which corresponds to a relaxation distance $L_2$
approximately ten times $l_0$) and $\varepsilon=0.09$, 0.07, and 0.065,
respectively.
For large values of the strain, the delaminated region propagates to all the film, however,
contrary to the previous case, there are large sectors in which the film remains attached to the substrate. The typical size
of these sectors is roughly determined by $L_2$, whereas the typical width of the delaminated stripes between sectors is order $L_1$. 
Note that the propagation
occurs via the formation and branching of undulated blisters, which at the end form a pattern of polygons bounded by undulated wrinkles.
When $\varepsilon$ is reduced, the branching of the undulated blisters reduces, giving rise to a pattern like that in Fig. \ref{f9}.
In a narrow
band of values of the applied strain (Fig. \ref{f10}), a single one dimensional delaminated region (a `telephone cord') may occur.
For lower values of $\varepsilon$ (not shown) the original blister is not able to grow at all.

\section{Conclusions}

I have presented results about the buckling of a thin film attached to a substrate. Buckling is induced by a mismatch in the relaxed 
configurations of film and substrate. For a fixed geometry of the delaminated region, namely a uniform stripe, the buckling patterns were
obtained as a function of the strain mismatch $\varepsilon$ and the Poisson ratio of the film $\nu$. The numerical results confirm the findings of previous
analytic treatments about the instability mode of the uniform Euler column: a undulated wrinkle for $\nu\gtrsim 0.25$, and a symmetric, `varicose'
pattern for $\nu\lesssim 0.25$. However, the present numerical analysis, which is not restricted to small distortions of the Euler column, shows that
the stability of the varicose pattern, in the region of experimental interest close to $\nu\simeq 1/3$, is restricted to a narrow interval of
$\varepsilon$ values, then limiting the possibility of observing this pattern experimentally.

In the second part of the paper I considered the interplay of buckling and delamination, and argued that a key factor determining the morphology
observed is the possibility or not of horizontal relaxation in the non-delaminated part of the film. For small horizontal relaxation, and large enough
$\varepsilon$, the film delaminates almost completely, generating a `brain-like' structure. When $\varepsilon$ is reduced, there is an abrupt transition
to a situation in which the delaminated region is not able to grow. 

For the case of large horizontal relaxation in the non-delaminated part of the film,
at large $\varepsilon$ the growing occurs via elongation and branching of
undulated delaminated regions. At the end a polygonal pattern delimited by undulated stripes is observed. 
The main characteristic of this pattern is the
existence of regions in which the film does not delaminate. When $\varepsilon$ is reduced, the tendency of the undulated stripes to 
branched is reduced, and eventually a regime where a single undulated stripe (a `telephone-cord') grows is observed. If $\varepsilon$ is reduced
further the  delamination stops completely. 

The simulations presented clarify the origin of the different morphologies observed experimentally. I want to emphasize in 
particular the importance of
horizontal relaxation in obtaining different kinds of patterns, most remarkably the necessity of horizontal relaxation to obtain the well known
telephone-cord pattern.

\section{Acknowledgments} 

This work was financially supported by Consejo Nacional de Investigaciones Cient\'{\i}ficas y T\'ecnicas (CONICET), Argentina, also under
grant PIP/5596 (CONICET).


\begin{thebibliography}{12}
\bibitem{ojo} To avoid confusion, note that $D$ is related with the bulk Young modulus $E$ of the material of the film
by $D=\frac{Ed^3}{12(1-\nu^2)}$, and  $B$ (which is the
proportionality between a given applied stress $\sigma$ in the plane of the film and the corresponding strain $\varepsilon$, namely
$\sigma=B\varepsilon$) can be written also as $B=E/(1-\nu)$.
\bibitem{nota2} The results of the simulations should be in principle dependent on the value of $k_b/(l_0^2k_s)$, even after
rescaling with the critical strain $\varepsilon_1$. This is equivalent to say that rigorously, we have three independent parameters in the problem:
$\nu$, $\varepsilon/\varepsilon_1$ (these two are used in the phase diagram of Fig. \ref{f2}), and $\varepsilon$ by itself. 
I have found that the explicit dependence in $\varepsilon$ is very weak, except in the rare cases in which 
$\varepsilon_1$ is very large.
\bibitem{otro} This estimation of the curvature is valid in the case when the applied $\varepsilon$ is much larger that the 
corresponding $\varepsilon_1$ calculated for a 
blister of typical size $L_1$. This condition is typically satisfied
in most experimental realizations.
\end{thebibliography}
\end{document}